%% file: main.tex
\documentclass[11pt]{article}
\usepackage{graphicx}
\usepackage[margin=1.25in]{geometry}
\usepackage[usenames,dvipsnames]{color}
\usepackage{cite} 
\usepackage{url}
\usepackage[colorlinks = true,
            linkcolor = blue,
            urlcolor  = blue,
            citecolor = blue,
            anchorcolor = blue]{hyperref}

\usepackage{lineno,hyperref,color}
\modulolinenumbers[5]
\usepackage{booktabs}
\usepackage{float}

\usepackage{appendix}
\usepackage{pdfpages}
\usepackage{hyperref}
\usepackage{geometry}                
\usepackage[parfill]{parskip}    
\usepackage{graphicx, subfigure}
\usepackage{amssymb}
\usepackage{amsmath}
\usepackage{epstopdf}
\usepackage{array}
\usepackage{lineno}
\usepackage{scrextend}

\def\Title#1{\begin{center} {\LARGE #1 } \end{center}}
\def\Author#1{\begin{center}{ \sc #1} \end{center}}
\def\Address#1{\begin{center}{ \it #1} \end{center}}

\newenvironment{Abstract}{\begin{quotation} \begin{center}
                       ABSTRACT
     \end{center}\bigskip  }{\end{quotation}}

\def\Acknowledgements{\bigskip  \bigskip \begin{center} \begin{large}
             \bf ACKNOWLEDGEMENTS \end{large}\end{center}}

\input workshopsymbols.tex
\input{defs}

\newcommand\snowmass{\begin{center}\rule[-0.2in]{\hsize}{0.01in}\\\rule{\hsize}{0.01in}\\
\vskip 0.1in Submitted to the  Proceedings of the US Community Study\\ 
on the Future of Particle Physics (Snowmass 2021)\\ 
\rule{\hsize}{0.01in}\\\rule[+0.2in]{\hsize}{0.01in} \end{center}}

\begin{document}


\medskip

\Title{Snowmass White Paper: Flavor Model Building}

\bigskip 

\Author{Wolfgang~Altmannshofer}
\Address{Department of Physics, University of California Santa Cruz, and
Santa Cruz Institute for Particle Physics, 1156 High St., Santa Cruz, CA 95064, USA}

\Author{Jure~Zupan}
\Address{Department of Physics, University of Cincinnati, Cincinnati, Ohio 45221, USA}

\medskip\medskip

\begin{Abstract}
\noindent In this white paper for the Snowmass process, we summarize the role flavor model building plays in the quest for new physics. We review approaches to address the non-generic flavor structure of the Standard Model and discuss how new physics models can be made compatible with the stringent constraints from flavor changing processes that indirectly probe very high scales. We also give an overview of the persistent anomalies in $B$ decays and the anomalous magnetic moment of the muon and some of their most popular new physics explanations.
\end{Abstract}

\snowmass

\def\thefootnote{\fnsymbol{footnote}}
\setcounter{footnote}{0}



\section{Executive Summary} \label{sec:exec}

Flavor model building is a prime example of experiment driven theoretical effort, which then in turn guides new experimental searches.
 Possibly the  most prominent historic example is the prediction of the charm quark based on the unexpectedly low rate of rare kaon decays and the subsequent discovery of the charm quark at colliders.
While still preliminary, the current experimental results on $b\to s \mu\mu$, $b\to c \tau \nu$ and $(g-2)_\mu$ might well be hints for possible new physics contributions, which would then mark the beginning of a new era in particle physics.
The experimental results jumpstarted a large model building undertaking that we highlight in some detail in a better part of this white paper. 
The main practical result is that, were these experimental hints for new physics to be confirmed, they would imply a new physical scale within reach of either the LHC or the next generation of colliders. The models that explain the $B$ physics anomalies include a number of new states such as a $Z'$, colorons, or leptoquarks, all of which can be efficiently searched for. 

Flavor model building is also a critical aspect of new physics model building that is motivated by longstanding questions in particle physics: the solution to the hierarchy problem, the origin of the Standard Model (SM) flavor structure, dark matter and the solution to the strong CP problem. 
In many instances, new physics model building becomes non-trivial precisely because flavor probes are highly sensitive to new physics, and the existing flavor constraints are very stringent. This is irrespective of whether or not the current flavor anomalies are a sign of new physics.  The new physics flavor problem is especially significant for UV completions of the SM that stabilize the electroweak scale and predict new states in the TeV regime with appreciable couplings to the SM particles. The solutions to the hierarchy problem of this type, such as composite Higgs models, models with extra dimensions, and models with low energy supersymmetry, all require a non-generic flavor structure that avoids flavor constraints.  The origin of such  non-generic new physics flavor structure is an open question, as is the origin of the hierarchies in the spectrum of the SM quarks and leptons, the so-called SM flavor puzzle. An important aspect of flavor model building is to construct mechanisms that address such open issues, implement them in new physics models, and derive phenomenological consequences, a selection of which are reviewed in this white paper.

\section{Flavor in the SM and Beyond} \label{sec:intro}

While the SM is exceptionally successful in describing particle physics phenomena, there is little doubt that new physics beyond the SM (BSM) exists. For one, dark matter has been discovered through its gravitational interactions, however, neither its mass nor the form of its interactions with the SM, if any, were yet uncovered.  Furthermore, arguments based on naturalness of the electro weak scale suggest that new physics degrees of freedom may exist at or below the TeV scale. These expectations are now in tension with null results from new physics searches at the LHC, which imply a significant mass gap between the electroweak scale and the scale of new physics that stabilizes the electroweak scale. 

Integrating out the heavy new physics states, the BSM effects can be described by non-renormalizable interactions of dimension $d>4$ that are suppressed by powers of the new physics scale $\Lambda$,  
\begin{equation}
\label{eq:L:EFT}
 \mathcal L_{\rm EFT}= \mathcal L^\text{SM}_\text{gauge} + \mathcal L^\text{SM}_\text{Higgs} + \sum_i \frac{1}{\Lambda^{d-4}} \mathcal C_i \mathcal O^{d>4}_i ~,
\end{equation}
where  the SM Lagrangian is just the first term in the expansion, i.e., the renormalizable part of an Effective Field Theory (EFT) expansion. Many searches for new physics can thus be performed without specifying the UV theory. Furthermore, the $d=4$ SM Lagrangian can be supplemented by new physics states that are light, but very weakly coupled. Most naturally these light new physics states are the (pseudo)-Nambu-Goldstone bosons that arise from spontaneously breaking of global symmetries. The most celebrated example is the QCD axion, whose existence would solve the strong CP puzzle. 

Flavor physics, i.e., the physics of processes in which quark or charged lepton flavors change,  plays a two-fold role (see also reviews \cite{Grossman:2017thq,Blanke:2017ohr,Gedalia:2010rj,Nir:2007xn,Kamenik:2017znu,Nierste:2009wg,Zupan:2019uoi,Gori:2019ybw,Silvestrini:2019sey}):
\begin{itemize}
 \item[(i)] First, the gauge sector of the SM exhibits a large accidental global flavor symmetry which is broken only by the Yukawa couplings of the Higgs to the SM quarks and leptons. 
The peculiar pattern of hierarchies in the Yukawa couplings is not explained in the SM, and calls for a dynamical new physics explanation. Since the SM Yukawa interactions are renormalizable, the scale at which new physics imprints the hierarchical structure can be arbitrarily high in principle.
\item[(ii)] Second, many of the higher dimension interactions in the EFT Lagrangian \eqref{eq:L:EFT} can, and are generically expected to, break the flavor symmetry of the SM gauge sector. Therefore, these interactions lead to quantifiable effects in low energy flavor changing processes. As long as the SM predictions and the experimental results for flavor changing processes are in agreement, this translates to strong indirect constraints on the new physics scale $\Lambda$.  Discrepancies in the flavor observables, on the other hand, could be the first indirect hints for new physics and establish a new scale in particle physics. 
\end{itemize}

Item (i) motivates new physics flavor model building based on a variety of mechanisms that can generate large hierarchies between fermion masses from generic $\mathcal O(1)$ input parameters. From practical point of view, the flavor models are particularly interesting, if they are anchored at experimentally accessible scales and make predictions that may be tested in not too distant future. Item (ii) relies on the interplay between the experimental precision program for measuring the CKM matrix elements as well as the searches for rare or forbidden flavor changing processes, and the corresponding theory effort to predict within the SM the relevant observables with comparable precision.

\section{Explaining the SM flavor hierarchies} \label{sec:SMflavor}

\begin{figure}[tb]
 \centering
 \includegraphics[width=1.0\textwidth]{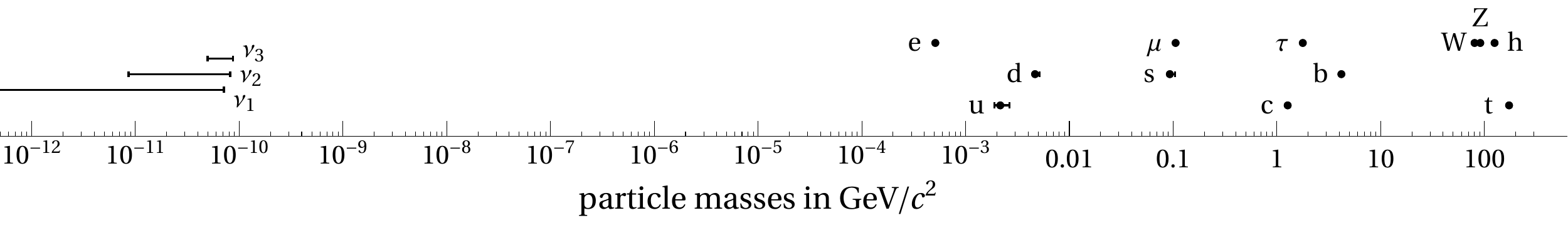} 
  \caption{Masses of the SM particles. The values for the masses of the Higgs, $W$, $Z$, quarks, and charged lepton masses are taken from~\cite{PDG}. The ranges for the neutrino masses are based on the known squared mass differences from neutrino oscillations~\cite{PDG} and the constraint on the sum of neutrino masses from Planck~\cite{Planck:2015fie}. The lightest neutrino could be massless.}
  \label{fig:masses}
\end{figure}

Figure~\ref{fig:masses} shows the masses of the SM particles~\cite{PDG}. The masses of the charged fermions span almost six orders of magnitude, from $0.5$\,MeV for the electron, to $173$\,GeV for the top quark. Quite strikingly, the top quark is the only SM fermion that has a mass of the order of the electroweak scale, $v=246$ GeV, while one would expect this to be the case for all the SM fermions, if the Yukawa couplings were generic, $Y_f\sim {\mathcal O}(1)$. The hierarchy of the charged fermion masses can be accommodated in the SM through hierarchical Yukawa couplings, $Y_e\ll\cdots \ll Y_t$,  but this pattern is not explained.

Neutrino masses are many orders of magnitude smaller than the masses of the charged fermions. While the absolute values of the neutrino masses are not known yet, at least two neutrinos need to be massive to accommodate neutrino oscillation data. The sum of the neutrino masses is also constrained by cosmological observations to be below few$\times 10^{-10}$~GeV~\cite{Planck:2015fie}, and thus at least six orders of magnitude lighter than the lightest charged lepton, the electron. This large gap is naturally explained, if the SM is viewed as an effective theory. In the minimal renormalizable SM, which does not contain right-handed neutrinos, the neutrinos are massless, while the neutrino masses then arise from the non-renormalizable Weinberg operator of dimension 5~\cite{Weinberg:1979sa,Minkowski:1977sc,Gell-Mann:1979vob,Yanagida:1980xy,Glashow:1979nm,Mohapatra:1979ia,Schechter:1980gr}.

The Yukawa couplings of the up-type and down-type quarks are not aligned in flavor space and the misalignment is parameterized by the CKM matrix. Similarly, the misalignment between the charged lepton Yukwas and the neutrino masses is parameterized by the PMNS matrix. The determination of the CKM and the PMNS matrix elements is a major collaborative effort between theory and experiment~\cite{Charles:2004jd,UTfit:2006vpt,deSalas:2017kay,Capozzi:2018ubv,Esteban:2018azc}. 
The absolute values of the CKM and PMNS matrix elements that are the result of this effort are approximately~\cite{PDG}
\begin{equation}
 |V_\text{CKM}| \sim \begin{pmatrix}
		              0.974 & 0.227 & 0.004 \\
                      0.226 & 0.973 & 0.041 \\
                      0.009 & 0.040 & 0.999 
                     \end{pmatrix} ~,\quad                      
 |V_\text{PMNS}| \sim \begin{pmatrix}
		              0.82 & 0.55 & 0.15 \\
                      0.32 & 0.60 & 0.74 \\
                      0.48 & 0.58 & 0.66 
                     \end{pmatrix} ~.\quad 
\end{equation}
While the CKM matrix clearly shows a hierarchical structure, the entries of the PMNS matrix are all within one order of magnitude.

The existence of the hierarchies between the various quark and charged lepton masses, the hierarchical structure of the CKM matrix, and the absence of visible hierarchies in the PMNS matrix is often referred to as the {\em SM flavor puzzle}. Several ideas have been put forward to explain the SM flavor puzzle using new dynamical degrees of freedom:

\paragraph{Horizontal flavor symmetries.} The hierarchical structure of the SM Yukawa matrices could be a result of a spontaneously broken (horizontal) symmetry under which fermions in different generations carry different charges, $[f_i]$. The simplest example are the Froggatt-Nielsen models of flavor \cite{Froggatt:1978nt}, where the SM gauge group is extended by a horizontal $U(1)_{\rm FN}$, and the matter field content by a set of vector-like fermions of mass $\sim M$. The SM fermions are charged under $U(1)_{\rm FN}$ in such a way that the SM Yukawa couplings, $Y_{f,ij} \bar f_{L,i} f_{R,j} H$, are forbidden since $x_{ij}\equiv [H]+[f_{R,j}]-[f_{L,i}]\ne 0$ for all $i,j$. The SM fermions thus couple to the Higgs through a chain of vectorlike fermions, giving hierarchical fermion masses $Y_{f,ij}\sim (\langle \varphi\rangle/M)^{|x_{ij}|}$ if the $U(1)_{\rm FN}$ breaking vacuum expectation value $\langle \varphi\rangle$ of the SM singlet scalar with $[\varphi]=-1$ is much smaller than the typical vectorlike fermion mass. Typically, the value $\langle \varphi\rangle/M\sim 0.2$ is used \cite{Leurer:1993gy,Leurer:1992wg,Ibanez:1994ig,Irges:1998ax,Fedele:2020fvh, Berezhiani:1989fp, Berezhiani:1990wn, Berezhiani:1990jj, Sakharov:1994pr}. The dynamical structure that explains the SM flavor puzzle brings in observable consequences. The tree level exchanges of a radial mode of $\varphi$ gives new physics contributions to meson mixing that are below present precision only if the mass of $|\varphi|$ is above $\sim 10^7$ GeV (in the so called clockwork limit \cite{Giudice:2016yja,Alonso:2018bcg} where $\varphi$ is not a dynamical field, the vectorlike fermions can be substantially lighter, with masses of a few TeV still allowed \cite{Buras:2011ph,Alonso:2018bcg}). The modulus, $\arg(\varphi)$, on the other hand is a pseudo-Nambu-Goldstone boson (pNGB), if $U(1)_{\rm FN}$ is a global symmetry. It thus can be light and be searched for directly. Since  $U(1)_{\rm FN}$ is anomalous under QCD the modulus $\arg(\varphi)$ can act as an axion and solve the strong CP problem (the so called {\em axiflavon} solution to the strong CP problem \cite{Calibbi:2016hwq,Ema:2016ops}, see also~\cite{Arias-Aragon:2017eww,Bonnefoy:2020llz,Cox:2021lii}). In the region of parameter space where the axiflavon can also explain the dark matter in the minimal model, it is within reach of the present and future rare kaon experiments \cite{Calibbi:2016hwq,Goudzovski:2022vbt}. Alternatively, if $U(1)_{\rm FN}$ is gauged, the $Z'$ can also be light, if the gauge coupling is small enough, and thus also be searched for directly \cite{Smolkovic:2019jow}.  The horizontal symmetry models can also be based on non-Abelian groups, most notably the $U(2)$ group \cite{Linster:2018avp,Barbieri:1996ww,Barbieri:1995uv,Pomarol:1995xc}, which is only minimally broken by the SM Yukawas \cite{Kagan:2009bn,Barbieri:2012uh,Barbieri:2011ci}. Going beyond horizontal symmetries, the unification of flavor and gauge symmetries has also been explored~\cite{Davighi:2022fer}.

\paragraph{Warped extra dimensions.} An interesting possibility that may point to a possible common origin of the smallness of the electroweak scale and the SM flavor structure is based on the idea of warped extra dimensions \cite{Randall:1999vf,Randall:1999ee}. In the Randall-Sundrum (RS) models of flavor, the geometry of space-time is five dimensional anti-de Sitter (AdS${}_5$), exhibiting a warped metric $ds^2=\exp(-2 k r_c |\phi|)\eta_{\mu\nu}dx^\mu dx^\nu -r_c^2 d\phi^2$, with $k$ the 5D curvature scale, $r_c$ the radius of compactification, and $\phi\in [-\pi, \pi]$ the coordinate along the fifth dimension.  A slice of AdS${}_5$ is truncated with two flat 4D boundaries, the Planck or UV and the TeV (IR) branes. The Higgs is localized on the TeV brane. The warp factor $\exp(-2 k r_c |\phi|)$ leads to different length scales along different 4D slices, which explains the apparent smallness of the Higgs vev from the 4D perspective, $\langle H\rangle_{\rm 4D}=\exp(-2 k r_c \pi) \langle H \rangle_5$, even though the 5D vev may be comparable to the Planck scale $\langle H \rangle_5\sim M_{\rm Pl}\sim 10^{19}$ GeV. For $k r_c \simeq 12$ one obtains $\langle H\rangle_{\rm 4D}\sim$ TeV. The fermion fields propagate in the bulk. The hierarchy between the SM charged fermion masses comes from exponentially suppressed overlaps of the fermion zero modes and the Higgs, where ${\mathcal O}(1)$ changes in the parameters of the 5D Lagrangian translate to exponential changes in these overlaps with the zero modes either localized near the UV or TeV brane \cite{Gherghetta:2000qt,Grossman:1999ra,Huber:2000ie}. While this set-up has a built in RS-GIM mechanism that suppresses too large FCNCs \cite{Agashe:2004ay,Agashe:2004cp}, the stringent constraints from precision flavor observables such as the indirect CP violating parameter $\epsilon_K$ in $K-\bar K$ mixing, the $B_{d(s)}-\bar B_{d(s)}$ and $D-\bar D$ mixing observables, as well as the rare decays such as $Br(\mu\to e \gamma)$, $Br(b \to s \gamma)$,..., already translate to bounds of ${\mathcal O}(20\text{\,TeV})$ on the masses of the first KK modes~\cite{Agashe:2004cp,Agashe:2003zs,Bauer:2009cf,Albrecht:2009xr,Blanke:2008zb,Casagrande:2010si,Beneke:2012ie,Malm:2015oda,Moch:2014ofa,Davoudiasl:2000my,Agashe:2006iy,Csaki:2010aj,Blanke:2012tv,Delaunay:2012cz},  two orders of magnitude above the weak scale and out of reach of the LHC. The RS constructions can thus be viewed merely as models of flavor, ignoring the little hierarchy problem. Alternatively, flavor symmetries or assumed nontrivial flavor structures for some of the couplings can be used to suppress further the FCNCs and thus lower the KK scale closer to the TeV scale \cite{Cacciapaglia:2007fw,Csaki:2008eh,DAmbrosio:2020ngh,Santiago:2008vq,Fitzpatrick:2007sa,Chen:2009gy,Csaki:2009wc,Agashe:2009tu}. Warped extra dimension models connecting the origin of flavor with the explanation of $B$ anomalies were developed in~\cite{Bordone:2017bld, Bordone:2018nbg, Fuentes-Martin:2022xnb}, see also discussion in Section \ref{sec:LQ}.

\paragraph{Partial compositeness.}
Partial compositeness \cite{Kaplan:1991dc,Agashe:2004rs,Contino:2003ve,Contino:2006qr} is a way of addressing the SM flavor puzzle in composite Higgs models, where the SM Higgs is a pNGB of a spontaneously broken global symmetry in the strongly coupled sector (see \cite{Panico:2015jxa} for a review and \cite{Carmona:2015ena,Barnard:2013zea,Ferretti:2013kya,Ferretti:2014qta,Vecchi:2015fma,Sannino:2016sfx,Cacciapaglia:2017cdi,Agugliaro:2019wtf,Cacciapaglia:2019dsq, Erdmenger:2020lvq, Erdmenger:2020flu} for concrete realizations). The SM fermions (and gauge bosons) are elementary particles that mix with their composite counterparts, the resonances in the strongly coupled sector, which carry the same SM quantum numbers. This is akin to photon-rho mixing at low energies. The lighter SM fermions are mostly elementary, while the heavier fermions, in particular the right-handed top are mostly composite. This hierarchical structure is assumed to be due to differing anomalous dimensions of the corresponding composite operators, which then translates to power suppressed fermion masses after renormalization group evolution from the UV to the weak scale even if one starts with an anarchic flavor structure in the UV. The hierarchy of mixings in partial compositeness then also leads to protection against excessive FCNCs, with the processes involving light SM fermions the least affected by the presence of the new composite sector. Even so, the kaon mixing constraints still require the scale of compositeness to be 10-20 TeV \cite{Csaki:2008zd}, significantly above the electroweak scale. In order for composite Higgs models to be the solutions to the hierarchy problem further flavor structure is therefore required \cite{Panico:2016ull,Redi:2011zi}.

\paragraph{Radiative fermion masses.}
Another class of models that address the SM flavor hierarchies is based on the idea of radiative fermion masses~\cite{Weinberg:1972ws}. In such models, the heavy fermions receive their mass by coupling to the Higgs at tree level. The light fermions, on the other hand, couple to the Higgs through loops of heavy new particles and therefore have strongly suppressed masses. 
This idea can be realized in many contexts, e.g., in supersymmetric models~\cite{Kagan:1989fp, Banks:1987iu, Arkani-Hamed:1996kxn, Borzumati:1999sp, Baumgart:2014jya, Altmannshofer:2014qha}, as well as in non-supersymmetric models~\cite{Balakrishna:1987qd, Balakrishna:1988ks, Dobrescu:2008sz, Baker:2020vkh}.
The ingredients that are common to all these models are new states and new sources of flavor violation, but there is a vast number of quantum numbers and interactions of the new states that can give viable scenarios. Models of radiative fermion masses predict that fermion masses of adjacent generations differ by a loop factor $\sim 1/16\pi^2 \sim 10^{-2}$,  in qualitative agreement with the observed spectrum of quarks and leptons.

\section{Flavor as the Probe of New Physics}

\subsection{Probing Heavy New Physics}

Flavor violating processes, in particular those based on flavor changing neutral currents (FCNC) have exquisite sensitivity to new sources of flavor and CP violation beyond the SM. The high new physics sensitivity has its origin in the minimal amount of flavor breaking that is present in the SM. 
In the SM, the only sources of flavor violation are the hierarchical Yukawa couplings of the Higgs, leading to quark FCNCs that are strongly suppressed by a loop factor and by small CKM matrix elements. As long as theoretical uncertainties in the SM predictions are under control, quark flavor violating processes can indirectly explore very high mass scales, in some cases far beyond the direct reach of collider experiments.
In the lepton sector, SM predictions for FCNCs are suppressed by the tiny neutrino masses and below any imaginable experimental sensitivities.  Electroweak contributions to electric dipole moments are also predicted to be strongly suppressed in the SM, several orders of magnitude below the current bounds. Charged lepton flavor violation and electric dipole moments are thus null tests of the SM. Any observation of such processes would be an unambiguous sign of new physics.

\begin{figure}[tb]
 \centering
 \includegraphics[width=.56\textwidth]{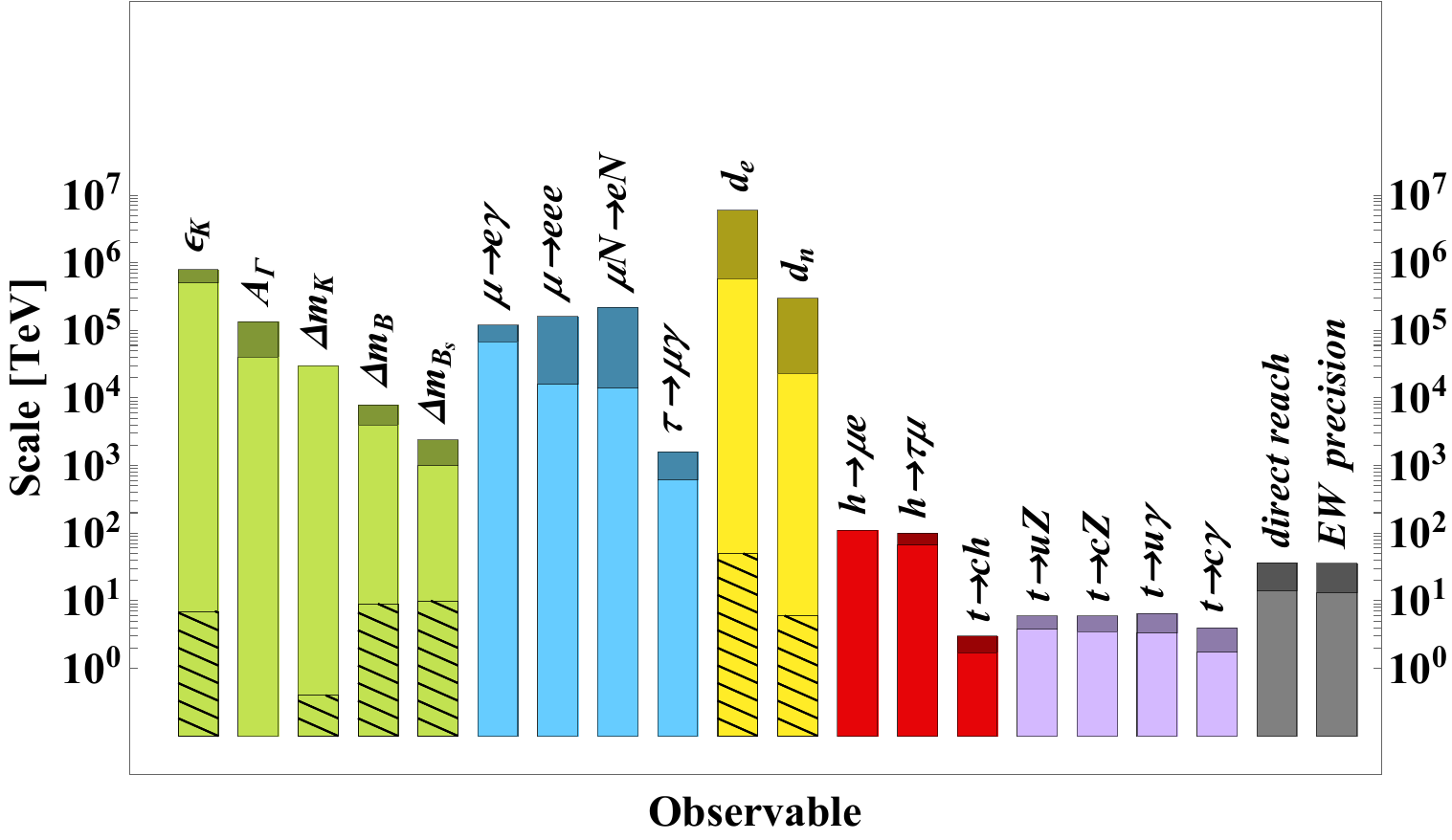}  ~~~
 \includegraphics[width=.40\textwidth]{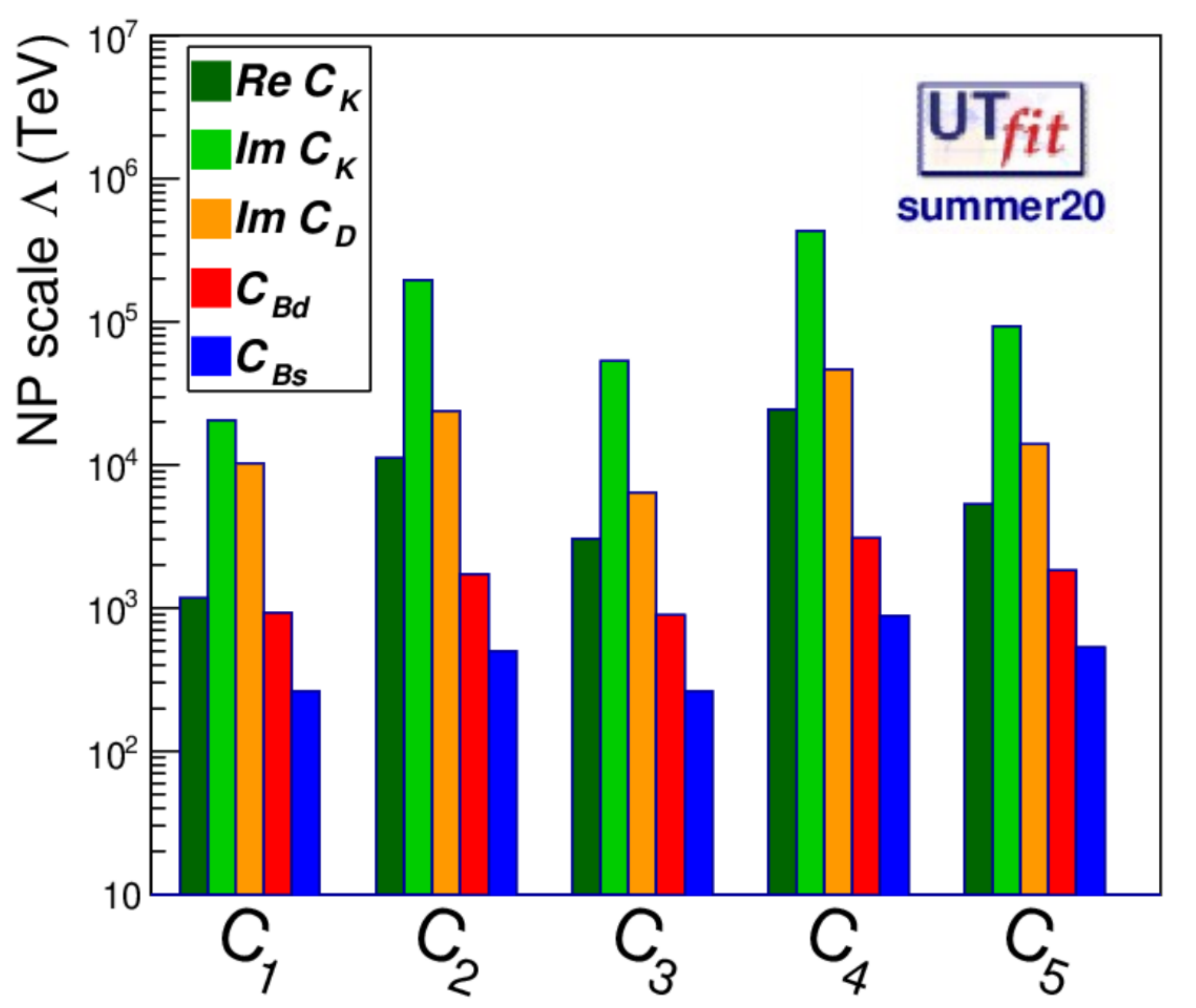}
  \caption{Comparison of the new physics reach for various flavor observables. Left: The generic new physics scale that can be probed with meson oscillations, lepton flavor violation, electric dipole moments, rare Higgs decays, rare top decays, direct searches, and electroweak precision observables. The light bars correspond to the current sensitivities, the dark bars to the expected future sensitivities. The hatched bars denote the sensitivities for scenarios with minimal flavor violation   (from~\cite{EuropeanStrategyforParticlePhysicsPreparatoryGroup:2019qin}).
  Right: Generic new physics scale that can be probed with meson mixing observables. The coefficients $C_{1\dots 5}$ correspond to different $\Delta F = 2$ four quark operators (from~\cite{UTfit}).  }
  \label{fig:NPscale}
\end{figure}

The high mass reach of several flavor changing processes is illustrated in the left plot of figure~\ref{fig:NPscale}. The plot assumes the presence of flavor violating dimension 6 interactions with $\mathcal O(1)$ Wilson coefficients. In that case, observables such as CP violation in kaon mixing, $\mu \to e \gamma$ transitions, and the electric dipole moment of the electron currently probe already exceptionally high scales $\sim \mathcal O(10^5~\text{TeV}) - \mathcal O(10^6~\text{TeV})$. Generically, flavor transitions from the second to the first generation are most strongly constrained.

In many cases more than one dimension 6 operator can contribute to a given flavor violating process. This is shown in the right plot of figure~\ref{fig:NPscale} for the example of meson mixing. The Wilson coefficients $C_1$ to $C_5$ correspond to flavor changing four quark operators that all have different chirality structure. Focusing for example on CP violation in kaon mixing (the light green bars), the scales that are probed for $\mathcal O(1)$ Wilson coefficients range from $\sim 2 \times 10^4$~TeV to $\sim 5 \times 10^5$~TeV, depending on the operator. Concrete new physics model may be described by a single operator or by several operators simultaneously. 

Arguments based on the naturalness of the electroweak scale suggest that there is new physics not far above the TeV scale. However, if there is such new physics, one would expect it to show up across many of the flavor changing processes mentioned above. The absence of deviations from the SM expectations in flavor observables that are most sensitive to new physics, assuming generic flavor violation, is known as the {\it new physics flavor puzzle}.

There are qualitatively different approaches to the new physics flavor puzzle:
\paragraph{Decoupling.} Deviations from the SM predictions are absent, if the new physics that introduces new sources of flavor violation is pushed to scales above the generic flavor constraints. The most prominent example of such a setup is mini-split supersymmetry (SUSY)~\cite{Arvanitaki:2012ps, Arkani-Hamed:2012fhg}. The spectrum of mini-split SUSY is such that the sfermions are a loop factor heavier than the gauginos. Flavor constraints, in particular kaon mixing, imply that the sfermion masses have to be approximately around the PeV scale while the gauginos, states that do not introduce any sources of flavor violation, can remain light, not far above a TeV~\cite{Altmannshofer:2013lfa,Isidori:2019pae}. (FCNC processes are loop suppressed in the minimal supersymmetric SM and therefore the scale that is constrained by meson mixing is not as high as in figure~\ref{fig:NPscale} but a loop factor smaller). Gauginos not far above a TeV are motivated by thermal WIMP dark matter, while the much heavier sfermions can effortlessly accommodate a Higgs mass of 125~GeV.

Mini-split SUSY and other models that avoid the new physics flavor puzzle by making (part of) the new physics spectrum heavy usually don't fully address the naturalness of the electroweak scale but only vastly reduce the amount of finetuning compared to the SM. The gap between the electroweak scale and the scale of flavored new physics might be explained by some form of anthropic selection, in line with the idea of ``frustrated naturalness''~\cite{Agashe:2022uih}.

\paragraph{Minimal Flavor Violation.} If there is new physics not far above the TeV scale, compatibility with the existing flavor constraints requires some form of flavor protection. The most extreme case corresponds to the absence of new sources of flavor violation. 
In any new physics extension of the SM, whether or not it is related to the solution of the SM flavor puzzle, there is a minimal amount of flavor violation that is inevitably present, namely due to the SM flavor Yukawas, $Y_f$. Even if the new physics is assumed to be completely flavor blind at the UV scale, radiative corrections proportional to $Y_f$ will introduce flavor breaking in the IR. If this is the only source of flavor breaking, the new physics sector is said to satisfy the Minimal Flavor Violation (MFV) hypothesis \cite{DAmbrosio:2002vsn,Hall:1990ac,Chivukula:1987py,Buras:2000dm,Ali:1999we,Gabrielli:1994ff}. Phenomenologically, the MFV is most often not needed in order to bypass flavor constraints, rather it is enough that new physics exhibits and approximate $U(2)^3$ structure that is broken by small spurions \cite{Kagan:2009bn,Barbieri:2011ci,Barbieri:2012uh}. Other minimally broken symmetries have also been considered~\cite{Arias-Aragon:2020bzy}.

\paragraph{Hierarchical new physics flavor couplings.}
Flavor symmetries that are broken in a controlled way can be used to suppress the amount of flavor violation in new physics models. More generally, the mechanisms discussed in section~\ref{sec:SMflavor} that can generate a hierarchical SM flavor structure, will typically generate hierarchies in the new physics flavor couplings as well and the most stringent flavor constraints may be avoided.

Interestingly, the simplest Froggatt-Nielsen models only lead to a very moderate suppression of new sources of flavor mixing. In particular, transitions between the first and second generation of quarks is only suppressed by one power of the spurion $\langle \phi \rangle/M \sim \lambda$. For example in a SUSY context, the strong constraints from kaon mixing imply that squarks need to be far above the electroweak scale \cite{Leurer:1993gy}. A similar situation arises in the simplest SUSY version of the flavor clockwork models~\cite{Altmannshofer:2021qwx}.  Much stronger suppression of new physics contributions to kaon mixing can be achieved in models with multiple $U(1)$ symmetries, for example in the so-called alignment models~\cite{Nir:1993mx}, or in the models with the non-abelian $U(2)$ flavor symmetries~\cite{Linster:2018avp, Barbieri:1996ww, Barbieri:1995uv, Pomarol:1995xc}. Models that strongly suppress transitions between the first and second generation may give detectable effects in $B$ physics.

\subsection{Higgs as the Flavor Probe}

In the SM, the Yukawa couplings of the Higgs to the fermions are the only sources of flavor violation. Therefore, the Higgs might be the window into understanding flavor, with the precision Higgs program at the LHC, and in the future also at a Higgs factory, able to provide valuable inputs. 

The SM predicts that at the tree level the Higgs couplings to fermions are flavor diagonal, CP conserving, and proportional to the fermion masses. Flavor changing couplings arise at the loop level, but are strongly suppressed below any foreseeable experimental sensitivity. Testing the SM predictions for the Higgs couplings is important to establish that the vacuum expectation value of the Higgs is the only source of SM fermion masses, and that the hierarchies in the fermion masses are indeed the result of hierarchical Yukawa couplings.

Existing measurements of Higgs production and decay rates have established that the couplings of the third generation quarks and leptons are SM-like with $\sim 10\%$ precision. The recent evidence for $h \to \mu^+\mu^-$~\cite{ATLAS:2020fzp,CMS:2020xwi} suggests that this is also the case for the coupling to muons. Confirming that the Higgs couplings to the remaining fermions are SM-like is very challenging~\cite{Kagan:2014ila,Altmannshofer:2015qra,Perez:2015aoa,Konig:2015qat,Perez:2015lra,Brivio:2015fxa,Bishara:2016jga} and alternative scenarios remain a viable option. Interesting scenarios are models in which the light fermions obtain their mass from a subdominant source of electroweak symmetry breaking~\cite{Altmannshofer:2015esa,Ghosh:2015gpa,Botella:2016krk,Das:1995df,Blechman:2010cs} (e.g. a second Higgs doublet or a technifermion condensate). Models of this type explain some of the hierarchies in the SM fermion mass spectrum not by hierarchical Yukawa couplings but instead by a hierarchy in sources of electroweak symmetry breaking. Two Higgs doublet models that implement this idea lead to characteristic collider signatures~\cite{Altmannshofer:2016zrn} (see also~\cite{Egana-Ugrinovic:2019dqu}).

Around the time of the Higgs discovery, once the Higgs mass was known, low energy flavor changing processes were used to constrain possible flavor violating couplings of the Higgs boson~\cite{Blankenburg:2012ex, Harnik:2012pb}. Very strong bounds can be derived from meson oscillations and $\mu \to e$ transitions. Barring tuned cancellations with unrelated contributions, most flavor changing decays of the Higgs were found to be constrained far below existing and expected experimental sensitivity. The exceptions are the Higgs decays involving tau, $h \to \tau \mu$ and $h \to \tau e$, for which the direct searches at the LHC are the most sensitive probes. Interestingly, the EFT arguments suggest that models without new sources of electro-weak symmetry breaking cannot give $h \to \tau \mu$ and $h \to \tau e$ rates at experimentally accessible level without violating the stringent bounds from rare tau decays~\cite{Altmannshofer:2015esa}. This provides continued motivation to search for the flavor violating Higgs decays, since these may well provide further insights into the origin of the fermion masses. 

\subsection{Flavor Transitions and Light New Physics}

Rare decays into a light new physics state, $X$, such as $K\to \pi X$ or $\mu\to e X$, are exquisite probes of new physics at high scales. Assuming completely anarchic couplings of $X$ to the SM the highest UV scale will typically be probed by the lightest initial state such that a decay to $X$ is still allowed. Taking as the example rare meson decays, $M=K,D,B$, the SM decay widths are power suppressed,  $\Gamma_M\propto m_M^5/m_W^4$. This then translates to parametrically enhanced sensitivities to decays involving light NP states.  In models where the NP particles couple to the SM via renormalizable interactions, and thus dimensionless couplings such as the mixing angle, $\theta$, between the Higgs and a light scalar $\varphi$, the NP decay width is ${\Gamma}(K\to\pi \varphi)\propto \theta^2 m_K$ and  consequently ${\mathcal B}(K\to \pi \varphi) \propto \theta^2 (m_W/m_K)^4$. This is to be compared with heavy meson decays: ${\mathcal B}(B\to K \varphi) \propto \theta^2 (m_W/m_B)^4$. For light new physics that couples to the SM through dimension-5 operators, such as axion-like particles (ALPs), the scaling changes to ${\mathcal B}(K\to \pi a) \propto  (m_W^2 /f_a m_K)^2$, to be compared to ${\mathcal B}(B\to K a) \propto (m_W^2 /f_a m_B)^2$, where $f_a$ is the ALP decay constant. These parametric enhancements  translate to a sensitivity to very high scales, much higher than in the case when rare decays are induced by off-shell new physics. It is also important to keep in mind that such naive dimensional analysis estimates can of course change, if the couplings of $X$ are not anarchic, but rather have a distinct flavor structure. In that case decays such as $B\to K X$ or $D\to \pi X$ can lead to the largest sensitivities even for very light $X$ masses. 

It is also interesting to translate the present and planned sensitivities of rare kaon, muon and $B$ meson factories to concrete models. Taking the QCD axion as a well motivated benchmark, the searches for $K\to \pi a$ and  $\mu \to e a$ decays, where $a$ escapes the detector, translate to bounds on the axion decay constant $f_a\gtrsim {\mathcal O} (10^{12})$ GeV \cite{MartinCamalich:2020dfe} and $f_a\gtrsim {\mathcal O} (10^{9})$ GeV \cite{Calibbi:2020jvd}, respectively, when assuming all flavor violating couplings are ${\mathcal O}(1)$. These bounds are more stringent than the astrophysics constraints, and so the improvements in searches for such rare decays could well lead to a discovery of the QCD axion. The scenarios where the QCD axion has flavor violating couplings include the possibility of  PQ symmetry being part of the horizontal flavor symmetry, in which case the solutions to the strong CP problem and the SM flavor puzzle would have a common origin. 

There are a number of other well motived light new physics models that could be probed by meson decays, a number of which have been recently discussed in detail in Ref. \cite{Goudzovski:2022vbt} for the case of rare kaon decays.  
For instance,  searching for $K\to \pi \varphi$  with two to three orders of magnitude larger datasets one could close the gap for Higgs-mixed scalar all the way to the big bang nucleosynthesis (BBN) floor. An improvement in sensitivity of ${\cal B}(K^+\to\ell^+N)$ by two orders of magnitude would start probing the minimal seesaw neutrino mass models for sterile neutrino masses in the ${\mathcal O}(100~{\rm MeV})$ regime. An order of magnitude improvement on ${\cal B}(K^+\to\mu^+\nu X_{\rm inv})$ would probe fully the preferred region for self-interacting neutrinos that may alleviate the Hubble tension. For heavier masses, $X$  would often decay inside the detector, leading to a number of signatures one could search for in rare $B$ decays, a possibility explored, e.g.,  for ALPs in \cite{Bauer:2021mvw,Ferber:2022rsf,Chakraborty:2021wda, Bertholet:2021hjl} and for inelastic dark matter in \cite{Filimonova:2022pkj,Kang:2021oes} (see also \cite{Dreyer:2021aqd}).

\section{Model Building for Flavor Anomalies} \label{sec:anomalies}
In the last several years a number of ``flavor anomalies''  created considerable excitement in the community. Discrepancies between SM predictions and experimental measurements are seen in $B$ decays (discussed in the remainder of this section) as well as in the anomalous magnetic moment of the muon (discussed in section~\ref{sec:g-2}).
If the new physics origin for these experimental anomalies could be established, it would have a transformative impact on the field. First and foremost, such an indirect sign of new physics would establish a new mass scale in particle physics. This scale could become the next target for direct exploration at future high-energy colliders. With sufficient energy, discoveries would, at least in principle, be guaranteed. 
Second, the couplings of the new physics constitute new sources of flavor violation beyond the SM Yukawa couplings. Existing low energy constraints suggest that such new physics couplings have a hierarchical flavor structure. This provides a new perspective on the Standard Model flavor puzzle and invites the construction of flavor models that link the structure of the SM and BSM sources of flavor violation.

\subsection{Overview of the $B$ Anomalies} \label{sec:anomalies_review}

Most prominent among the flavor anomalies are the hints for lepton flavor universality (LFU) violation in the charged current $b \to c \ell \nu$ transitions~\cite{BaBar:2012obs, LHCb:2015gmp, Belle:2015qfa, Belle:2017ilt, LHCb:2017rln, Belle:2019rba} and in the neutral current $b \to s \ell \ell$ transitions~\cite{LHCb:2017avl, LHCb:2019efc, LHCb:2021trn, LHCb:2021lvy}. In the SM, lepton flavor universality corresponds to an accidental approximate symmetry that is only broken by the small lepton Yukawa couplings. Up to kinematical effects, the decay rates of $b \to c \ell \nu$ and $b \to s \ell \ell$ decays with different lepton flavors are expected to exhibit lepton universality, and ratios of the corresponding branching ratios (observables like $R_{D^{(*)}} = \text{BR}(B\to D^{(*)}\tau\nu)/\text{BR}(B\to D^{(*)}\ell\nu)$ and $R_{K^{(*)}} = \text{BR}(B \to K^{(*)} \mu\mu)/\text{BR}(B \to K^{(*)} ee)$) can be robustly predicted in the SM. 

The observed deviations from the SM predictions of $R_{D^{(*)}}$ have a combined significance in the range of $3.1\sigma$ to $3.6\sigma$, depending on how error correlations are treated~\cite{HFLAV:2019otj, Bernlochner:2021vlv}. The combined significance of the observed non-standard values of $R_{K^{(*)}}$ exceeds $4\sigma$~\cite{Altmannshofer:2021qrr, Isidori:2021vtc}. The $R_{K^{(*)}}$ anomalies are accompanied by several additional anomalies observed in neutral current muonic $b \to s \mu \mu$ decays. Deviations from the SM predictions are for example observed in the $B \to K^* \mu\mu$ angular distributions~\cite{LHCb:2020lmf, LHCb:2020gog} as well as in the absolute $B \to K^* \mu\mu$, $B \to K \mu\mu$, and $B_s \to \phi \mu\mu$ branching ratios~\cite{LHCb:2014cxe, LHCb:2016ykl, LHCb:2021zwz}. While the corresponding SM predictions are under lesser control compared to the theoretically exceptionally clean LFU ratios, the pattern of deviations is remarkably consistent with the new physics explanations of $R_{K^{(*)}}$.

\subsection{Model Independent Considerations} \label{sec:model_independent}

Model independently, new physics effects in the $b \to s \ell \ell$ and $b \to c \ell \nu$ decays can be described by an effective Hamiltonian
\begin{equation}
 \mathcal H_\text{eff} = \mathcal H_\text{eff}^\text{SM} - \frac{4 G_F}{\sqrt{2}} V_{tb} V_{ts}^* \frac{\alpha_\text{em}}{4\pi} \sum_i C_i^\text{NC} Q_i^\text{NC} + \frac{4 G_F}{\sqrt{2}} V_{cb} \sum_i C_i^\text{CC} Q_i^\text{CC} ~
\end{equation}
where $Q_i^\text{NC}$, $Q_i^\text{CC}$ are dimension six operators that mediate the $b \to s \ell \ell$ and $b \to c \ell \nu$ transitions, respectively, and $C_i^\text{NC}$, $C_i^\text{CC}$ are the corresponding Wilson coefficients. With the chosen normalization factors, the relevant SM Wilson coefficients are of $\mathcal O(1)$. New physics contributions to the Wilson coefficients of $\mathcal O(1)$ correspond to generic new physics scales of $\Lambda_\text{NC} = |\frac{4 G_F}{\sqrt{2}} V_{tb} V_{ts}^* \frac{\alpha_\text{em}}{4\pi}|^{-1/2} \simeq 35$~TeV and $\Lambda_\text{CC} = |\frac{4 G_F}{\sqrt{2}} V_{cb}|^{-1/2} \simeq 0.9$~TeV.
Such a setup can capture all new physics that is heavy compared to the $b$ hadrons. In the presence of new light degrees of freedom (e.g. sterile neutrinos or light di-lepton resonances) dedicated studies are required (see e.g. \cite{Sala:2017ihs, Altmannshofer:2017bsz, Datta:2017ezo, Darme:2021qzw, Greljo:2021npi, Crivellin:2022obd, Asadi:2018wea, Greljo:2018ogz}). 

New physics contributions to the Wilson coefficients will generically modify a lrge set of observables (total rates as well as kinematic and angular distributions) in several decay modes. The available experimental information is then combined with theory predictions (that involve lattice QCD input on hadronic matrix elements) in global fits to identify possible new physics explanations of the anomalies. The following effective operators turn out to be the leading candidates for an explanation of the anomalies~\cite{Shi:2019gxi, Murgui:2019czp, Geng:2021nhg, Altmannshofer:2021qrr, Hurth:2021nsi, Alguero:2021anc, Ciuchini:2021smi}
\begin{equation}
 C_9^{bs\mu\mu} (\bar s \gamma_\alpha P_L b) (\bar \mu \gamma^\alpha \mu) ~,\quad C_{10}^{bs\mu\mu} (\bar s \gamma_\alpha P_L b) (\bar \mu \gamma^\alpha \gamma_5 \mu) ~,\quad  C_V^{bc\tau\nu} (\bar c \gamma_\alpha P_L b) (\bar \tau \gamma^\alpha P_L \nu_\tau) ~.
\end{equation}
The $b\to s \ell \ell$ decays point to new physics contributions $C_9^{bs\mu\mu} \simeq -0.8$, or $C_9^{bs\mu\mu} = - C_{10}^{bs\mu\mu} \simeq -0.4$. The $b\to c \ell \nu$ data is best described by $C_V^{bc\tau\nu} \simeq 0.07$. The mere existence of consistent new physics explanations is a non-trivial result. The preferred values for the Wilson coefficients point to generic new physics scales of few TeV in the case of the charged current decays, to few 10's of TeV in the case of the neutral current decays~\cite{Altmannshofer:2017yso,DiLuzio:2017chi}.

The anomalies can be tested in a model independent way at colliders. Explanations of the $b \to s \ell \ell$ anomalies are expected to affect the high energy tails of di-lepton spectra at the LHC and future hadron colliders~\cite{Greljo:2017vvb} and lead to enhanced $\mu^+ \mu^- \to bs$ production at a high energy muon collider~\cite{Altmannshofer:2022xri}. Explanations of the $b \to c \tau \nu$ anomalies are expected to give non-standard mono-tau production at the LHC~\cite{Altmannshofer:2017poe, Greljo:2018tzh, Marzocca:2020ueu}.

The results from the global fits form the basis for the constructions of new physics models to explain the anomalies.

\subsection{Models with $Z^\prime$ Bosons} \label{sec:Zprime}

Many $Z^\prime$ scenarios have been put forward as possible explanations of the neutral current $b \to s \ell \ell$ anomalies.
However, they can not explain the anomalies in the charged current $b \to c \ell \nu$ decays.
One popular class of models is based on gauging the difference of muon-number and tau-number, $L_\mu - L_\tau$~\cite{He:1990pn,He:1991qd}. This gauge group is anomaly free already given the SM particle content and leads to vectorial couplings of the $Z'$ to muons (and taus). Once this $Z'$ model is augmented by physics that introduces flavor violating $Z'$ couplings to quarks, it can explain the observed discrepancies in $b \to s \ell \ell$ decays~\cite{Altmannshofer:2014cfa, Crivellin:2015mga}. 
A simple construction involves heavy vector-like fermions that are charged under $L_\mu - L_\tau$ and that can mix with the SM quarks~\cite{Fox:2011qd, Altmannshofer:2014cfa}.

Besides $L_\mu - L_\tau$, various other combinations of gauged flavor symmetries have been used to construct $Z'$ models that can address the $b \to s \ell \ell$ anomalies (for a few examples see~\cite{Crivellin:2015lwa, Celis:2015ara, Fuyuto:2015gmk, Bonilla:2017lsq, Bhatia:2017tgo, Alonso:2017uky, King:2017anf, Allanach:2018lvl, Altmannshofer:2019xda, Bhatia:2021eco, Greljo:2021xmg, Greljo:2021npi}).
Also scenarios where the $Z'$ couples to both quarks and leptons indirectly have been considered in the context of the rare $B$ decay anomalies.
The $Z'$ can also find a natural home in models with partial compositeness~\cite{Niehoff:2015bfa,Sannino:2017utc,Carmona:2017fsn,Chung:2021ekz}.

In models with $Z^\prime$ bosons, meson mixing puts strong constraints on the flavor changing couplings. One therefore finds {\it upper} bounds on the $Z^\prime$ masses that are typically around several TeV, possibly in reach of the LHC or future colliders~\cite{Greljo:2017vvb, Allanach:2017bta, Abdullah:2017oqj, Kohda:2018xbc, Allanach:2019mfl, Huang:2021nkl}. Many models predict additional heavy states (e.g. the vector-like fermions in \cite{Altmannshofer:2014cfa}) that are clear targets for future colliders. 

Different $Z^\prime$ models often make characteristic predictions for other low energy flavor processes. 
For example, the generic expectation in models with partial compositeness is that the couplings are strongest to the third generation, reflecting the mass hierarchy of the SM fermions that is related to their degree of compositeness. Correspondingly, such models often predict large enhancements of the $b \to s \tau \tau$ decays and also sizeable rates for lepton flavor violating decays. 
In contrast, $L_\mu - L_\tau$ models predict the absence of lepton flavor violating decays at detectable levels. Moreover, these models predict effects that are equal in size but opposite in sign for $b \to s \tau \tau$ and $b \to s \mu \mu$ decays. Given the current data, this corresponds to modest $\sim 20\%$ enhancements of the $\tau\tau$ modes. This motivates future Tera-Z machines which would have unique sensitivities to the experimentally challenging $b \to s \tau \tau$ decays~\cite{Kamenik:2017ghi,Li:2020bvr}.

\subsection{Models with Leptoquarks} \label{sec:LQ}

Leptoquarks are very popular explanations of the flavor anomalies. In contrast to the $Z^\prime$ bosons, leptoquarks do not contribute to meson mixing at tree level. Contributions arise first at 1-loop and constraints from meson mixing therefore leave ample room in leptoquark parameter space. Among the full set of leptoquarks that can have renormalizable couplings to the SM quarks and leptons, several can provide explanations of the neutral current $b \to s \ell \ell$ anomalies or the charged current $b \to c \ell \nu$ anomalies~\cite{Hiller:2014yaa, Alonso:2015sja, Bauer:2015knc, Fajfer:2015ycq, Barbieri:2015yvd, Bhattacharya:2016mcc, Hiller:2016kry,  Crivellin:2017zlb, Angelescu:2018tyl, Popov:2019tyc, Cornella:2019hct, Perez:2022ouu}. 

Leptoquarks are contained in many different BSM scenarios. Both scalar and vector leptoquarks could be part of the composite spectrum of a strongly coupled sector above the TeV scale~\cite{Gripaios:2014tna, Barbieri:2016las, Marzocca:2018wcf}. One of the scalar leptoquarks that can explain $R_{D^{(*)}}$ can be identified with the right-handed sbottom in the Minimal Supersymmetric Standard Model with R-parity violation~\cite{Deshpande:2016yrv, Das:2017kfo, Altmannshofer:2017poe, Earl:2018snx, Trifinopoulos:2018rna}. Vector leptoquarks can be the gauge bosons of an enlarged gauge group that is broken above the TeV scale~\cite{DiLuzio:2017vat, Bordone:2017bld, Barbieri:2017tuq, Calibbi:2017qbu, Greljo:2018tuh, Blanke:2018sro, Balaji:2018zna, Fornal:2018dqn}. Also scalar leptoquarks can arise in models of quark-lepton unification~\cite{FileviezPerez:2013zmv}.

Existing direct searches for leptoquarks at the LHC probe leptoquark masses up to $\sim 1.5$~TeV. Most leptoquark models that explain $R_{D^{(*)}}$ predict deviations in di-tau production at the LHC. The HL-LHC should be able able to cover the preferred parameter space of those models. Leptoquarks that explain the $b \to s \ell \ell$ anomalies can be much heavier, outside the reach of the LHC~\cite{Allanach:2017bta, Hiller:2018wbv}. They would lead to discoverable effects in di-jet production at a high energy muon collider~\cite{Huang:2021biu, Asadi:2021gah}.

Interestingly, there is a single leptoquark that can address both the $b \to s \ell \ell$ and the $b \to c \ell \nu$ anomalies simultaneously: the vector leptoquark $U_1$ with the SM quantum numbers $(3,1,2/3)$. This leptoquark can be identified as one of the gauge bosons of the Pati-Salam (PS) gauge group. A considerable amount of flavor model building in recent years has been motivated by the $U_1$ explanation of the flavor anomalies and possible embeddings of the $U_1$ leptoquark into UV complete setups. One scenario that has emerged as particularly promising has quarks and leptons interacting via generation specific PS gauge groups, the so-called PS$^3$ models~\cite{Bordone:2017bld, Bordone:2018nbg, Fuentes-Martin:2022xnb}. The ``deunification'' in flavor space can be used to generate hierarchies in the leptoquark couplings and might very well be also related to the SM flavor puzzle (for a $U(2)$ based model see \cite{Barbieri:2019zdz}).

Most realistic leptoquark models predict many additional states (colorons, $Z^\prime$ bosons, ...) at  scales that are accessible with colliders.
They also predict characteristic effects in many low energy flavor observables, for example strongly enhanced rates for $b \to s \tau \tau$ decays, lepton flavor violating $B$ decays, or modest enhancements in $b \to s \nu\bar\nu$ decays, all signatures in reach of LCHb or Belle II. 

\section{Anomalous magnetic moment of the muon} \label{sec:g-2}

There is a long-standing discrepancy between the SM prediction and the experimental results on the anomalous magnetic moment of the muon $a_\mu = \frac{1}{2}(g-2)_\mu$.
The combination, $a_\mu^{\rm avg}$, of the measurements by the Muon $g-2 $ collaboration at Fermilab~\cite{Abi:2021gix,Albahri:2021kmg,Albahri:2021ixb} and previously at BNL~\cite{Muong-2:2006rrc}, differs by $4.2\sigma$~\cite{Abi:2021gix} from  the consensus SM prediction, $a^{\rm SM}_\mu$~\cite{Aoyama:2020ynm},  $\Delta a_\mu
=   a^{\rm avg}_\mu - a^{\rm SM}_\mu
=   \left( 251 \pm 59 \right) \times 10^{-11}$ (the  BMW collaboration prediction using lattice QCD, on the other hand, is consistent with experiment at $1.6\,\sigma$~\cite{Borsanyi:2020mff}, but awaits confirmation by other lattice QCD groups). Interpreting this discrepancy as a hint for new physics, there are two classes of new physics models that can explain $\Delta a_\mu$, depending on whether or not the required chirality flip  occurs on the internal new physics line in the loop (see also surveys of models in \cite{Lindner:2016bgg,Jegerlehner:2009ry,Crivellin:2018qmi,Athron:2021iuf}). If the chirality flip occurs on the muon line, this introduces a suppression by the muon mass and thus the new physics running in the loop need to be light. A prime example of a model of this type is a contribution to $(g-2)_\mu$ from a light $Z'$ \cite{Pospelov:2008zw}. Requiring that this is a gauge boson from an anomaly free $U(1)_X$ with the minimal particle field content, it  can lead to the shift in the magnetic moment of the muon large enough to explain the measured $\Delta a_\mu$, without being excluded by other constraints, mainly if the  $Z'$ is in the 100 MeV mass range \cite{Altmannshofer:2019zhy,Greljo:2021npi}. The constraints on the $Z'$ depend on how it couples to the other SM fermions. The gauged $L_\mu-L_\tau$ in general faces the least severe constraints \cite{He:1990pn,He:1991qd,Altmannshofer:2014cfa,Altmannshofer:2019zhy,Greljo:2021npi,Alonso-Alvarez:2021ktn,Cen:2021iwv,Coloma:2020gfv,Heeck:2018nzc,Biswas:2021dan}. Other examples include flavor violating ALPs \cite{Bauer:2019gfk}, from photonic couplings of ALPs \cite{Davoudiasl:2018fbb,Marciano:2016yhf,Buen-Abad:2021fwq}, photonic and leptonic couplings of ALPs \cite{Buttazzo:2020vfs}, or from ALP with a dark photon \cite{Ge:2021cjz}.

If the NP contributions to $(g-2)_\mu$ receive a chirality flip from the internal line, such contributions are not suppressed by the muon mass, and the new physics states can be more massive, in the several TeV range. Such models require at least two new physics fields, in order to have the large chirality flip possible on the internal line. Examples of models of this type include: muophilic dark matter running in the loop \cite{Jana:2020joi,Chowdhury:2021tnm,Calibbi:2018rzv,Borah:2021khc,Acuna:2021rbg,Kowalska:2020zve,Chen:2020tfr,Kawamura:2020qxo,Calibbi:2019bay}, contributions from ALPs coupling to heavy vectorlike leptons \cite{Brdar:2021pla}, anomalous $Z'$ \cite{Capdevilla:2021kcf,Capdevilla:2021rwo},  singlet scalars \cite{Capdevilla:2021kcf,Capdevilla:2021rwo}, low energy supersymmetry \cite{Baum:2021qzx,Berger:2008cq,Stockinger:2006zn,Feng:2011aa,Cho:2011rk,Altmannshofer:2021hfu}, extended Higgs sectors \cite{Escribano:2021css,Arcadi:2021zdk,Bharadwaj:2021tgp,Dermisek:2021ajd}, radiative models for charged fermion or neutrino masses \cite{Baker:2021yli,Greljo:2021npi,Chen:2020jvl,Calibbi:2020emz}. There may also be a relation with $B$ anomalies \cite{Belanger:2015nma,Arcadi:2021cwg,Perez:2021ddi,Heeck:2022znj,Arnan:2019uhr,Allanach:2015gkd,Navarro:2021sfb}, while implications for Higgs physics were discussed in \cite{Crivellin:2021rbq,Endo:2020tkb,Fajfer:2021cxa,Harnik:2012pb,Dermisek:2013gta,Davoudiasl:2012ig}.

The challenge for the new physics models explaining the $(g-2)_\mu$ anomaly is the absence of any such new physics hints in lepton-flavor-violating transitions, such as $\mu\to e\gamma$ and $\mu \to 3 e$. For a generic flavor structure these give bounds on the new physics scale that is much higher than what is required for $(g-2)_\mu$. Phrasing these constraints in terms of the effective new physics suppression scale for the dimension 5 dipole moment operators, $\mathcal{L}_{\rm eff} \supset - { e \,v}\, \bar \ell^i_{LL} \sigma^{\mu \nu} \ell^j_{RR } F_{\mu \nu}/{(4 \pi \Lambda_{ij})^2} + {\rm h.c.}$, where $v  ={246}${GeV} is the electroweak vev, and $i,j$ generational indices,  the new physics scale required to explain the $(g-2)_\mu$ anomaly  is $\Lambda_{22} \simeq {15}$\,{TeV}, while the absence of $\mu \to e \gamma$ implies $\Lambda_{12 (21)}  \gtrsim {3600}$\,{TeV}~\cite{MEG:2016leq}. The flavor violating transitions therefore need to be significantly suppressed, either by ad-hoc flavor alignment of new physics couplings, or through the use of symmetries, such as the $U(1)_X$ gauge symmetry in the case of light $Z'$ explanation for $(g-2)_\mu$ \cite{Greljo:2021npi}.

\section{Conclusions}

Flavor physics plays a dual role. Firstly, the observed SM flavor structure calls for a dynamical explanation and motivates new physics model building. Secondly, the rare flavor changing processes are sensitive probes of new physics. The reach depends on the assumed flavor structure of new physics couplings and spans scales from just above the electroweak scale, if minimal flavor violation is assumed, all the way to scales as high as $10^{12}$ GeV, probed by the searches for the flavor violating QCD axion with anarchic couplings. Intriguingly, there are hints for possible deviations from the SM expectations in the measurements of  $b\to s \mu\mu$ and $b\to c\tau \nu$ transitions, and  in $(g-2)_\mu$. 
If these flavor anomalies are indeed signs of new physics, this would imply a very bright and phenomenologically rich future ahead of us. Many discoveries at the high energy and high intensity frontiers would in that case be expected in the not too distant future.

\Acknowledgements
The research of W.A. is supported by the U.S. Department of Energy grant number DE-SC0010107. W.A. also acknowledges support by the Munich Institute for Astro- and Particle Physics (MIAPP) which is funded by the Deutsche Forschungsgemeinschaft (DFG, German Research Foundation) under Germany's Excellence Strategy – EXC-2094-390783311. J.Z. acknowledges support in part by the DOE grant de-sc0011784 and NSF OAC-2103889.

\bibliographystyle{JHEP}
\bibliography{theBIB}

\end{document}

%% file: workshopsymbols.tex
\def\Acknowledgements{\bigskip  \bigskip \begin{center} \begin{large}
             \bf ACKNOWLEDGEMENTS \end{large}\end{center}}

\def\beq{\begin{equation}}
\def\eeq#1{\label{#1}\end{equation}}
\def\eeqn{\end{equation}}

\newenvironment{Eqnarray}%
   {\arraycolsep 0.14em\begin{eqnarray}}{\end{eqnarray}}
\def\beqa{\begin{Eqnarray}}
\def\eeqa#1{\label{#1}\end{Eqnarray}}
\def\eeqan{\end{Eqnarray}}

\let\bar=\overbar

\def\lsim{\mathrel{\raise.3ex\hbox{$<$\kern-.75em\lower1ex\hbox{$\sim$}}}}
\def\gsim{\mathrel{\raise.3ex\hbox{$>$\kern-.75em\lower1ex\hbox{$\sim$}}}}

\def\del{\partial}
\def\Dslash{\not{\hbox{\kern-4pt $D$}}}
\def\dslash{\not{\hbox{\kern-2pt $\del$}}}
\def\pslash{\not{\hbox{\kern-2pt $p$}}}
\def\ETmiss{\not{\hbox{\kern-4pt $E$}}_T}

\def\Dlr{\mathrel{\raise1.5ex\hbox{$\leftrightarrow$\kern-1em\lower1.5ex\hbox{$D$}}}}

\def\MSB{{\bar{M \kern -2pt S}}}
\def\msb{{\bar{\scriptsize M \kern -1pt S}}}

\def\drb{{\bar{\scriptsize D \kern -1pt R}}}

